\documentclass[12pt]{article}

\usepackage[dvips]{color}
\usepackage{multicol}
\usepackage{amsmath,color,fancybox,graphics,graphpap,rotating}
\definecolor{white}{rgb}{1,1,1}
\definecolor{yellow}{rgb}{0.95,0.75,0.1}
\definecolor{red}{rgb}{0.5,0,0}
\definecolor{green}{rgb}{0,1,0}
\definecolor{bgcolor}{rgb}{0.94,0.91,0.78}

\definecolor{lblue}{rgb}{0,0.8,1}
\definecolor{dblue}{rgb}{0,0,.6}
\definecolor{dgreen}{rgb}{0,0.3,0}
\definecolor{lila}{rgb}{0.8,0,0.8}
\definecolor{violet}{rgb}{1,0,1}
\definecolor{grey}{rgb}{0.3,0.3,0.3}
\definecolor{turquoise}{rgb}{0,.9608,1}

\definecolor{contoura}{rgb}{0,0,1}
\definecolor{contourb}{rgb}{0,1,1}
\definecolor{contourc}{rgb}{0,1,0}
\definecolor{contourd}{rgb}{0.95,0.75,0.1}
\definecolor{contoure}{rgb}{1,0,0}
\definecolor{contourf}{rgb}{1,0,1}

\def\lsim{\raise0.3ex\hbox{$\;<$\kern-0.75em\raise-1.1ex\hbox{$\sim\;$}}}
\def\gsim{\raise0.3ex\hbox{$\;>$\kern-0.75em\raise-1.1ex\hbox{$\sim\;$}}}
\newcommand{\nee}{\nonumber \end{eqnarray}} 
 
\usepackage{amsmath,color,fancybox,graphics,graphpap,rotating}
\newcommand{\be}{\begin{eqnarray}}
\newcommand{\ben}{\begin{eqnarray}\nonumber}
\newcommand{\ee}{\end{eqnarray}}

\definecolor{lred}{rgb}{1,0.3,0.3}
\definecolor{red}{rgb}{0.5,0,0}
\definecolor{dblue}{rgb}{0,0,.6}

\def\ARAA{{ Ann. Rev. Astron. \& Astrophys.} }
\def\ApJ{{ Astrophys. J.} }

\def\AA{{ Astron. \& Astroph.} }

\def\MNRAS{{ Month. Not. Roy. Astr. Soc.} }
\def\Nature{{ Nature} }

\def\PRD{{ Phys. Rev.} {\bf D} }

\begin{document}

\title 
{Degeneracy Breakdown as a Source of Supernovae Ia}

\author{
L. Clavelli\footnote{Louis.Clavelli@Tufts.edu,\,lclavell@bama.ua.edu}\\
Dept. of Physics and Astronomy, Tufts University, Medford MA 02155\\
Dept. of Physics and Astronomy, Univ. of Alabama, Tuscaloosa AL 35487}
\date{Sept 3, 2016}
\maketitle

\begin{abstract}
We pursue the investigation of a model for sub-Chandrasekhar supernovae Ia explosions (SNIa) 
in which the
energy stored in the Pauli tower is released to trigger a nuclear deflagration.
The simplest physical model for such a degeneracy breakdown is a phase transition
to an exactly supersymmetric state in which the scalar partners of protons, neutrons, and leptons become degenerate with the familiar fermions of our world as in the supersymmetric standard model with susy breaking parameters relaxed to zero.  We focus on the ability of the susy phase transition model to fit the total SNIa rate as well as the 
delay time distribution of SNIa after the birth of a progenitor white dwarf.  We also study the
ejected mass distribution and its correlation with delay time.  F‌inally, we discuss the expected SNIa remnant in the form of a black hole of Jupiter mass or lower and the prospects for detecting such remnants.  
\end{abstract}


\section{Introduction{\label{Intro}}}

   In the 1930's S. Chandrasekhar \cite{Chandra} famously showed that electron degeneracy pressure would make certain stars now known as white dwarfs classically stable up to about 1.4 solar masses. Slightly below this mass spontaneous nuclear fusion would erupt to destabilize the star.
In 1973 it was proposed \cite{Whelan-Iben},\cite{Webbink} that mass accretion onto white dwarfs 
from a binary partner could cause this Chandrasekhar mass to be approached from below at which point nuclear fusion would take over leading to an explosion which could be identified as a type Ia supernova. The clear evidence of fusion by-products expected from a
Carbon or Oxygen White Dwarf progenitor supports the Whelan-Iben idea.  In addition, the prediction that supernovae occur at the Chandrasekhar mass suggests
a possible explanation for the crucial uniformity of the events.     

Now, more than four decades later, a quantitative understanding of the 
explosion and the supernova properties has been elusive in this now standard model.  In addition, there is now strong evidence for a sub-Chandrasekhar 
component.  


Within the standard model for SNIa there is a distinction between the
single degenerate (SD) scenario in which there is a main sequence star donating
matter to the degenerate white dwarf and the double degenerate (DD) scenario in
which a second degenerate white dwarf donates to a primary, more massive white dwarf
progenitor.  If both mechanisms are operative the supernovae uniformity is more difficult to understand.  Both mechanisms are subject to a host of problems as
discussed in a number of old and recent reviews for example \cite{Hillebrandt}, \cite{Maoz-Mannucci}.  In particular both mechanisms
when faced with observational constraints greatly underestimate the SNIa rate.
Historically, when a phenomenon resists explanation in a standard model for
multiple decades, the resolution of the puzzle often requires some radical new     
physics input.

Up until some five years ago the DD scenario was strongly disfavored and
the obstacles to a satisfying theory based on this scenario were summarized as follows in the still cogent 2000 Review by Hillebrandt and Niemeyer \cite{Hillebrandt}:

 ``Besides the lack of convincing direct observational evidence for sufficiently many appropriate binary systems, the homogeneity of `typical' SNe Ia may be an argument against this class of progenitors. It is not easy to see how the merging of two white dwarfs of (likely) different mass, composition, and angular momentum with different impact parameters, etc, will always lead to the same burning conditions and, therefore, the production of a nearly equal amount of $^{56} Ni.$''   

On the other hand, the SD scenario, although conceptually easier to envision,
comes with its own set of puzzles.  For instance there is no evidence
 of a binary partner remnant and no evidence of
 significant absorbtion by the partner nor of significant asymmetry in the explosion as might be expected from a planar binary system. 

 The dilemma was heightened in 2010 with the observation that accretion in 
 the SD scenario would necessarily be accompanied by significant X-ray emission
 and observed X-ray activity was far below what would be required if the entire
 SNIa rate was to be understood as accretion onto white dwarfs several tenths of 
 a solar mass below the Chandrasekhar mass.
 Quantitatively, whereas an average accretion rate of $100\, M_\odot/Gyr$ would be
 required to bring $1.2\, M_\odot$ dwarfs to the Chandrasekhar mass at a sufficient
 rate, no more than one to two percent of this rate is consistent with
 observations \cite{Bogdan-Gilfanov}\,\cite{DiStefano}.  Recently, an analogous constraint on the
 the DD scenario has been proposed based on the absence of expected polarization
 in supernova light \cite{Bulla}.

  The purpose of this article is to refine the analysis of the susy phase
  transition model for SNIa \cite{Biermann-Clavelli}.  In this model every white  
  dwarf,
  whether in a binary system or not, has a characteristic mass-dependent lifetime.
  Most white dwarfs have a lifetime longer than the current age of the universe 
  but those in a certain sub-Chandrasekhar mass range can have lifetimes in the 
  tenths of a Gigayear range.  Unless made explicit, masses in this article are
  defined relative to the solar mass, $M_\odot$. 

 The primary parameters of this
  model were a critical matter density, $\rho_c$, and a minimum lifetime, 
  $\tau_0$.  
  We generalize the model via a third free parameter, $b_0$, to allow the 
  possibility of a phase transition suppression at high pressure analogous to that
  observed in a superheated liquid.  Finally, we assume, as a fourth parameter, 
  an average white dwarf accretion rate parametrized by some $c_0$
  which causes white dwarfs to increase in mass up to that at which a
  significant fraction of the white dwarfs undergo the supernova phase transition. 
  This parameter is required to be consistent with the above-mentioned limits on 
  the accretion in binary systems of a white dwarf with a main sequence partner.
  
  The number of these parameters is comparable to the minimum number in 
  the standard model (mean and width of of the double mass distributions
  and mean and width of the accretion rate distribution or initial separation
  distribution).  Clearly no prediction
  of the supernova rate and delay time distributions can be made without
  knowing or assuming initial state distributions and then they can be tested 
  against present or future observations.  At present no known binary systems are
  clearly supernova candidates.  In the phase transition model, the supernova 
  rates and delay time distribution are proportional to known white dwarf mass
  distributions.   

   In recent years, relatively precise analyses have been made of the delay time
 distribution of SNIa.  It is found that most of these events occur within a few
 tenths of a Gigayear after the birth of the progenitor white dwarf with only
 about $1\%$ occuring after Gigayear (Gyr) delay times.  This seems surprising in
 view of common multi-Gyr stable orbits of binary objects.
 Prediction of the delay time distribution in the standard accretion models
 would seem to depend sensitively on unknown binary mass distributions,  
 unknown orbital parameter distributions, and unknown accretion rates.  Unless 
 the corresponding distributions are surprisingly narrow, the supernovae uniformity
 is again puzzling.

  In the framework of the phase transition model we are able to quantitatively
  fit the total SNIa rate and the rates in three recently observed delay time bins, 
  The ejected mass distribution is also predicted and shows a sharp peaking.
  
\begin{figure}[ht]
\centering
\includegraphics[scale=0.65]{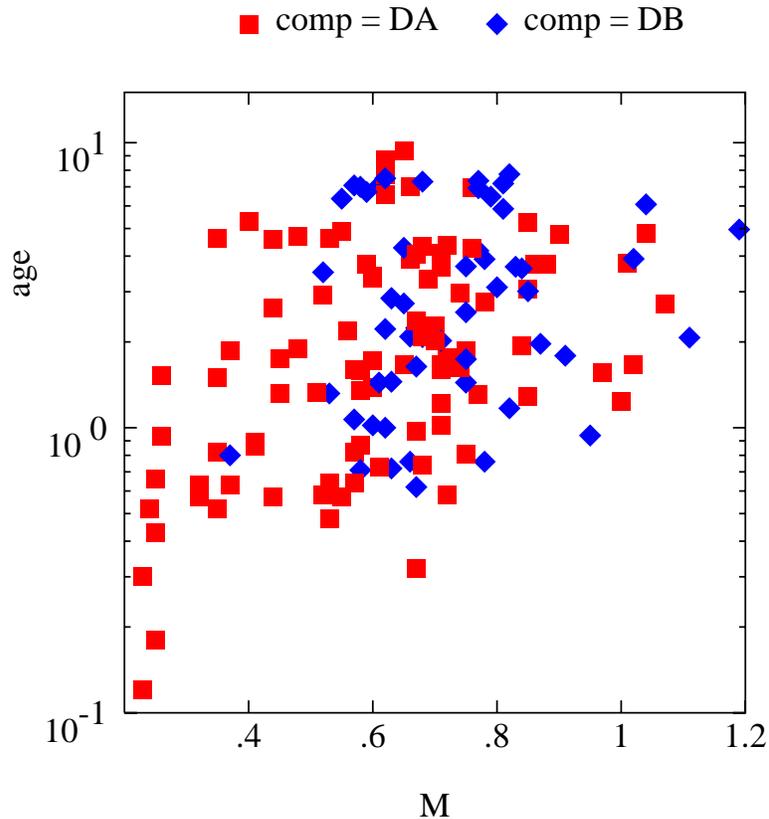}
\caption{The age in Gyr versus mass in a clean sample of old white dwarfs.} 
\label{CoolWDs} 
\end{figure}

The most recent high statistics data release from the Sloan Digital Sky Survey
(SDSS) contains a clean sub-sample in which the mass of the white dwarfs
and their age from birth as a white dwarf
are reasonably well determined \cite{Bergeron}.  This sample shown in fig.\ref{CoolWDs}
contains 95 DA type white 
dwarfs which have a thin hydrogen atmosphere (about $10^{-5}$ of the total mass)
and 55 DB type white dwarfs which have a similarly thin helium atmosphere.
A clear dip in the distribution near a mass of about $0.45$ Solar was noted.

In addition, fig.\ref{CoolWDs} suggests a dip in the DB distribution at
a higher mass as well as an extension to higher masses compared to the DA white
dwarfs.  This could suggest that the DB dwarfs have arisen from an
earlier DA phase due to an enhanced accretion leading to hydrogen to helium fusion on the surface.  In this paper, however, we restrict our attention to the DA
white dwarfs.  

  The organization of this paper is as follows.  In Section \,\ref{WDdists}, we start with 
  the mass distribution of hot white dwarfs which should approximate the 
  distribution at birth.  We then compare the mass distribution of the old white 
  dwarfs with known ages \cite{Bergeron}. This allows us to 
  estimate the average accretion rate. 

   In Section \,\ref{phasetrans} we briefly review the theory of the phase transition to exact susy
   from \cite{Biermann-Clavelli} and discuss the possibility of extending the 
   model to include a transition-suppressing pressure term in the action.  The 
   analog would be the suppression of the boiling transition in a liquid under
   high pressure.  Conservation of degrees of freedom in this model requires the
   existence of a broken susy in our world although the masses in this phase can 
   be quite high consistent both with their non-observation in current accelerator 
   searches and with susy grand unification theory.

   In Section \,\ref{Carlo} we describe a four parameter monte carlo and discuss how  
   the parameters are constrained by $1 \sigma$ fits to the 
   observed delay time distribution.

   In Section \,\ref{massdists} we discuss the ejected mass distribution and its
correlation with delay times.

   In Section \,\ref{remnants} we discuss the predicted supernova remnants and their   
   eluding of current searches as well as some consequences for future searches. 

   Section \,\ref{summary} is reserved for a summary and discussion of results.

\section{White dwarf mass distributions and accretion\label{WDdists}}

\begin{figure}[ht]
\centering
\includegraphics[scale=0.65]{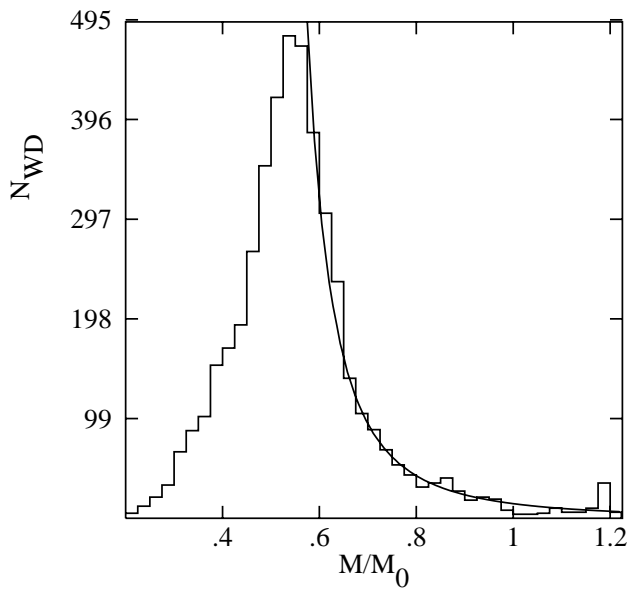}
\caption{The SLOAN sample of hot ($T_{eff}> 12000K$) DA white dwarfs \cite{Madej04},\cite{Madej07} compared to the theoretical fit from eq.\,\ref{WDIMF1} in the high
mass tail.}
\label{hotDAdwarfs} 
\end{figure}

The white dwarf initial mass function $F(m(0))$ above the peak at $0.6 \,m_\odot$
is obtained from the Salpeter initial mass function for a main sequence star followed by an initial to final mass function which linearly relates the white dwarf production rate at birth mass, m(0), to that of the parent star.
The relation
\be
    \frac{dN}{dm(0)} = a_0 F(m(0)) = a_0 ((m(0) - 0.478)/0.09028)^{-2.35}  \qquad m(0)>0.62 M_\odot 
\label{WDIMF1}
\ee
gives a good fit to the decreasing part of the hot white dwarf mass distribution as shown in figure \,\ref{hotDAdwarfs}, \cite{Madej04}.  The curve shown corresponds to
$a_0 = 1000$ for the observed sample in that plot.  

We will take this distribution to represent that of white dwarfs at birth (t=0)
At later times we will assume an average mass increase due to accretion corresponding to the form
\be
    m(t) = m(0) + c_0 t
\label{accretion}
\ee 
with $c_0$ being a free parameter in the monte carlo.  Of course, particular stars
could accrete at a lower or higher rate and we assume that an average
accretion rate is a useful first approximation.  
     
We could phenomenologically extend the fit down to $0.35 M_\odot$
with the form
\be 
    \frac{dN}{dm(0)} =F(m(0)) = 0.474 a_0 / (1 + (m(0) - .62) ^ 2 / .048) ^ 2   \qquad 0.35 < m(0) < 0.62\qquad .
\label{WDIMF2}
\ee

However, in our monte carlo we find that no white dwarf with a birth mass less
than 0.8 results in a supernova in less than a Hubble time so we can ignore eq.\ref{WDIMF2}.

The SLOAN data shown in fig.\,\ref{hotDAdwarfs} indicate that a fraction $f=0.292$
have masses above $0.6\, M_\odot$.
We normalize to a sample of $10^{10}$ white dwarfs, similar to the Milky Way, 
following the hot white dwarf mass distribution shown in fig.\ref{hotDAdwarfs}.  We assume
that the birth rate of white dwarfs has been constant for $12.8$ Gyr and is proportional to this hot white dwarf mass distribution.  That is, with a new
$a_0$ normalized to this prototype galaxy,
\be
    \frac{d^2 N_{WD}}{dt dM(0)} = a_0 \,F(M(0))
\label{birthrate}
\ee
with
\be
  f \cdot 10^{10} = a_0 \,12.8 Gyr\, \int_{0.6} ^{1.4} \,dM(0)\, F(M(0))
\ee
so that
\be
     a_0 = 5.48\, {WD/yr/galaxy} \quad .
\label{a_0}
\ee

Here we have assumed that the accretion rate is small enough to be ignored as is
the white dwarf depletion due to supernovae.  In the double degenerate scenario
the supernova rate is proportional to the probability to produce two white dwarfs
of total mass near $1.4\, M_\odot$ which is presumably significantly less than the probability
from eq.\ref{WDIMF1} to produce a single white dwarf of Chandrasekhar mass.

As a rough measure of the accretion rate
we take the $95$ cool white dwarfs with measured masses and ages from ref.\,\cite{Bergeron}.  We assume that some fraction $f_0$ are isolated and therefore do not 
accrete and that the the others have accreted at an average rate $c_0$.  Under these
assumptions Eq.\,\ref{accretion} then statistically defines a birth mass. The peak of the hot white dwarf mass distribution comes from solar mass stars that have a
long lifetime on the main sequence.  The high mass tail could, however, reveal the
scale of accretion leading to the cool white dwarf spectrum.  We scan over values 
of $f_0$ and $c_0$ seeking the least $\chi^2$ in comparison with fig.\,\ref{hotDAdwarfs} and eq.\,\ref{WDIMF1}.  In table\,\ref{table0} we plot the 
best fit predicted numbers at birth $N(M(0))$ to compare with
eq.\,\ref{WDIMF1} normalized at a mass of $0.7$.

\begin{table}[ht]
\begin{center}
\begin{tabular}{|l||c||c|}\hline
        M    &\, N(M(0))\,  &\, eq.\,\ref{WDIMF1} \cr\hline
       0.5   &\,  10	    \,  &\,  11.8 \cr\hline
       0.6   &\,  20	    \,  &\,  19.7 \cr\hline
       0.7   &\,  26        \,  &\,  26.0 \cr\hline
       0.8   &\,  6	    \,  &\,  10.05 \cr\hline
       0.9   &\,  3	    \,  &\,  5.75 \cr\hline
       1.0   &\,  5	    \,  &\,  3.49 \cr\hline
       1.1   &\,  1	    \,  &\,  2.31 \cr\hline
       1.2   &\,  0	    \,  &\,  1.63 \cr\hline
       1.3   &\,  0	    \,  &\,  1.20 \cr\hline
\end{tabular}      
\caption{Best fit numbers of hot white dwarfs $N(M(0))$ at mass $M$
compared to the observed numbers from eq.\,\ref{WDIMF1} assuming a fraction
$f_0$ is isolated (non-accreting) and that the others accrete with average 
rate $c_0$.}
\label{table0}
\end{center}
\end{table}

The resulting best fit estimates are
\be
     f_0 = 0.34 \qquad c_0 = 0.004\,M_\odot \quad .
\ee 

The average accretion rate is independently determined below in
a monte carlo of the phase transition model. 

Above a birthmass of $1.1$ the numbers in table\,\ref{table0} predicted from the cool white dwarf data
are significantly lower than the initial mass function which could be taken as an indication that, due to supernovae, a significant
number of white dwarfs originally at mass of $1.1$ or above do not survive 
cooling to low temperatures.

\section{Degeneracy breakdown in dense matter\label{phasetrans}}

We briefly review here the model of \cite{Biermann-Clavelli} for the release of 
the energy stored in dense matter due to the Pauli exclusion principle.
Further details can be found in that reference. 

Earlier papers on the susy phase transition in the context of gamma ray
bursts were refs.\cite{Clav-Karatheodoris}, \cite{Clav-Perevalova}, \cite{growth},
\cite{gravcollapse}.

   In the string landscape picture, the universe can exist in a large number
of local minima of the vacuum energy.  The transition probability per unit space time volume from one local minimum to another is given in the thin wall approximation by the vacuum decay
formula \cite{Coleman}
\be 
     \frac{d^3 P}{dt d^3 r} = A e^{-B(r)}
\ee
where the vacuum decay action, $B$, is proportional to the inverse cube of the 
difference between the vacuum energies.  One would expect that, in the presence of matter,
this action is proportional to the inverse cube of the total energy density difference.  In
dense stars the vacuum energy density is negligible compared to the matter density difference. As discussed in ref.\cite{Biermann-Clavelli}, the matter density
difference is equal to the Pauli energy. 
In the Fermi gas model, this energy stored in the Pauli towers is
proportional to the matter density, $\rho(r)$, so the action $B(r)$, in first approximation, can be written
\be
    B(r) = (\frac{\rho_c}{\rho(r)})^3
\label{B}
\ee
where the critical density $\rho_c$, treated as a free parameter, is related to the surface tension of the bubble wall between the two phases.  In a white dwarf star
with Chandrasekhar density profile $\rho(r)$, the transition probability per unit time is
\be
    \frac{dP}{dt} = \frac{1}{\tau_0 V_0} \int d^3 r\, e^{-B(r)} = \frac{1}{\tau}
\quad .
\label{dPdt}
\ee
Without loss of generality one can choose $V_0$ such that the free parameter,
$\tau_0$, is the minimum lifetime over all white dwarfs.
Thus there are two primary free parameters, a critical density $\rho_c$ and a
minimum lifetime, $\tau_0$.

The Chandrasekhar density profile is defined by the balance between the
gravitational pressure gradient
\be
     \frac{dP_G}{dr} = - \rho(r) G M(r)/r^2
\ee
and the degeneracy pressure gradient
\be
  \frac{dP_D}{dr} = \frac{ab}{3} \rho^{-2/3} \frac{x^4}{1+x^2} \frac{d\rho}{dr}
\ee
where a is proportional to the classical electron energy density
\be
    a = \frac{{m_e}^4 c^5}{3 \pi^2 \hbar^3} = \frac{m_e c^2}{3 \pi^2 {r_0}^3}\quad ,
\ee
$r_0$ being the classical radius of the electron, 
\be 
     b = \frac{\pi \hbar}{m_e c}(\frac{3}{2 \pi m_N})^{1/3} \quad ,
\ee
with $m_N$ being the nucleon mass and
\be
     x = b \rho^{1/3} \quad .
\ee

Using Chandrasekhar's calculation one finds that for low mass dwarfs, the
density is small leading to a large lifetime.  For masses approaching the
Chandrasekhar mass ($\approx 1.4 M_\odot$) the volume approaches zero leading
again to a large lifetime.  Consequently, some white dwarfs very close to the
Chandrasekhar mass could be long-lived while slightly lighter dwarfs would 
rapidly decay.  The same effect could come from a pressure term in the action
as discussed below.  Thus the lifetime, $\tau$, 
is expected to be a quasi-parabolic function of white dwarf mass.  

Of course one might expect that there are sub-leading corrections to the 
Coleman-DeLuccia formula.  This formula gives the probability per unit
space time volume to nucleate the first critical bubble of true vacuum.  If this 
probability is large the initial nucleation is followed rapidly by further
bubble creation.

One could also ask whether there are transition suppressing pressure terms in the
action increasing with density such as in the case of a super-heated liquid.
Thus, with this analogy, one could allow the possibility that the action is enhanced at high density parametrized by a $b_0$:
\be
    B(r) = (\frac{\rho_c}{\rho(r)})^3 + b_0 (\rho(r)/\rho_c)^{4/3} \quad .
\ee

\begin{figure}[ht]
\centering
\includegraphics[scale=0.65]{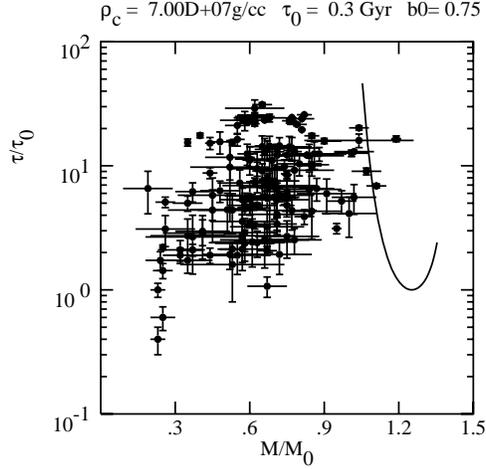}
\caption{The lifetime relative to the minimum $\tau_0$ as a function of mass
relative to solar mass for critical density $\rho_c = 7\cdot 10^7\,$ g/cm$^3$, 
with $\tau_0 = 0.3\,$ Gyr and $b_0 = 0$.  
For comparison, the average density of a typical white dwarf is about
$M_\odot/{R_{Earth}}^3 = 7.66 \cdot 10^6\,$ g/cm$^3$.  
Also shown are the 95  
DA white dwarfs in the clean sample of ref.\cite{Bergeron} with ages
relative to $\tau_0$ plotted on the y axis.
}
\label{SNIL}
\end{figure}

For a particular choice of parameters, the lifetime as a function of 
white dwarf mass is shown in fig.\ref{SNIL}.

Neglecting accretion, white dwarfs are born at zero age and grow vertically
in the plot of fig.\ref{SNIL}.  The survival probability to age $t$ is then
\be
    P_s = e^{-t/\tau} \quad .
\label{survpro}
\ee
In the absence of accretion and with the given parameter choice one would expect to find no old white dwarfs in the quasi-parabolic region. 

\section{The four parameter monte carlo\label{Carlo}}

In fitting the observational data on SNIa we scan over values of the four 
parameters $\rho_c, \tau_0, b_0,$ and $c_0$.  We find that the latter two parameters
can shift the solution space somewhat but are not critical to finding a solution.
In the presence of accretion, the transition probability per unit time becomes
\be
    \frac{dP}{dt} = \frac{1}{\tau_0 V_0} \int d^3 r\, e^{-B(r)} = \frac{1}{\tau(M)}
\ee
where the white dwarf mass $M$ is related to the birth mass by eq.\ref{accretion}.
The survival probability to age t then becomes
\be
    P_s(t) = e^{-\int_{0}^{t} dt^\prime /\tau(M(t^\prime))}\quad .
\ee

In the absence of accretion, $M$ is constant and the survival probability reduces to
eq.\,\ref{survpro}.  Old white dwarfs in the parabolic region should be those
that are rapidly accreting which might be susceptible to observational test.

\begin{figure}[ht]
\centering
\includegraphics[scale=0.65]{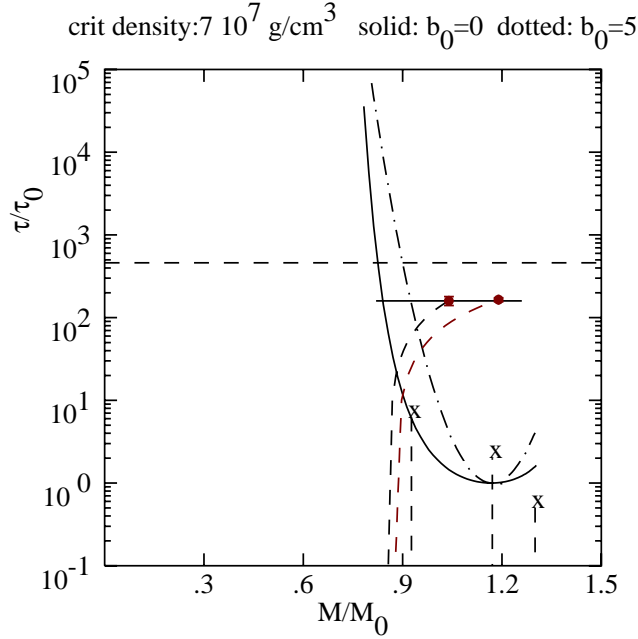}
\caption{Paths to supernova without accretion (vertical lines) and with accretion
(curved lines). The $\tau(M)/\tau_0$ curve is approximated by the best fit parabola.
The horizontal dashed line corresponds to the age of the universe divided by
$\tau_0 = 0.03\,\mathrm{Gyr}$.} 
\label{ageing}
\end{figure}

\begin{figure}[ht]
\centering 
\includegraphics[scale=0.65]{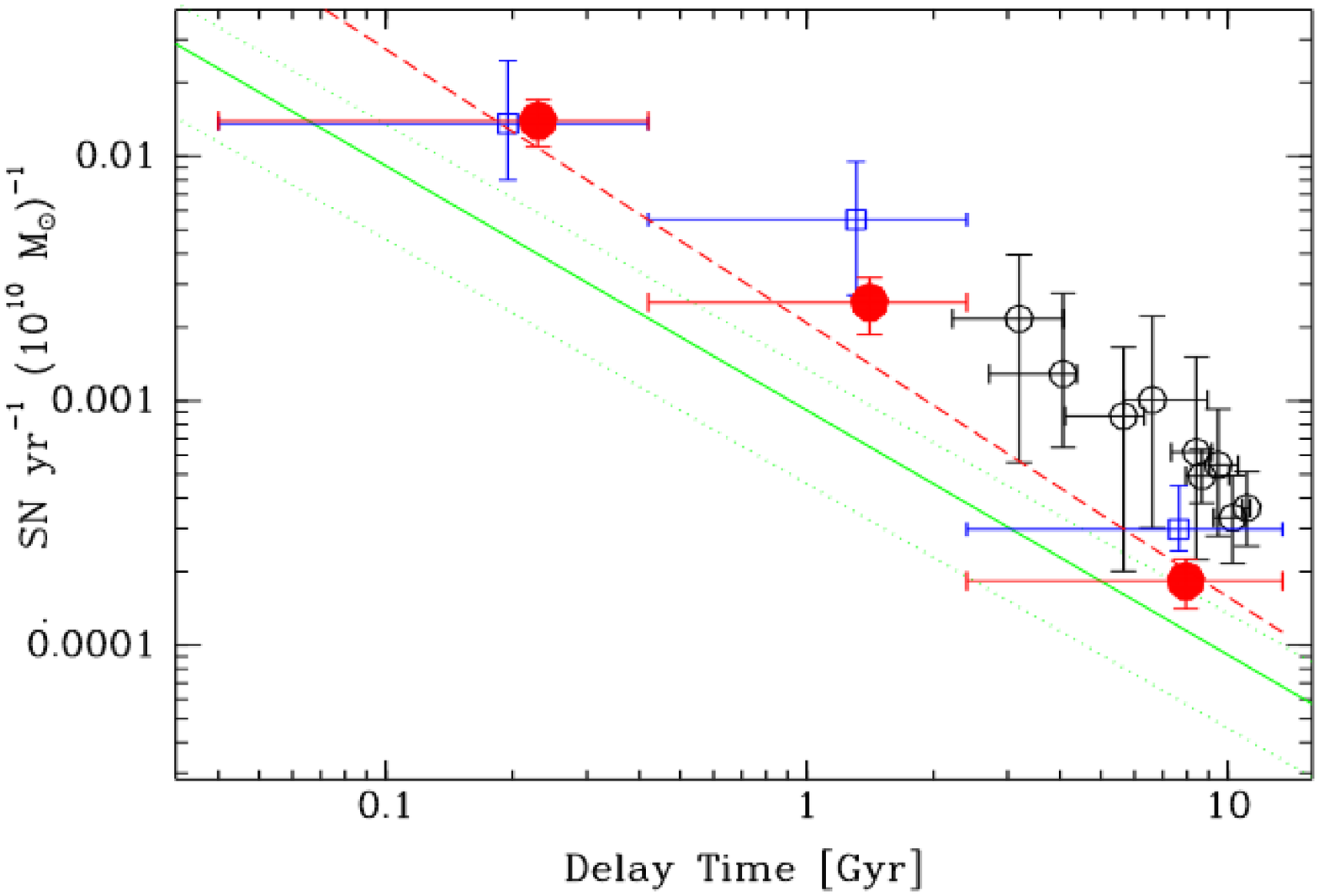}
\caption{The SNIa rate as a function of delay time from \cite{Maoz12}. See text.  The key point is that
most supernova occur within $0.4\,\mathrm{Gyr}$ of the white dwarf birth but the delay
time distribution extends into the multi-Gyr range.} 
\label{DTdist}
\end{figure}

 In the case of no accretion (e.g. solitary white dwarfs) the stars age vertically
in fig.\ref{ageing}. The probability of surviving until reaching the $\tau(M)$ curve is then $1/e$.  With the indicated critical density, the two shown high mass DA dwarfs from the Bergeron et al. sample \cite{Bergeron} are unlikely to survive to the observed age without accretion but can have a high survival probability if they are accreting and age along the curved paths. Alternatively, the critical density parameter can be increased which results in the parabola moving to the right.
Neglecting the fall-off of $F(M(0))$ in the region of $M \approx 1.2$, the greatest supernova rate comes from the
minimum of the $\tau(M)$ curve. The two white dwarf masses above and below the
minimum have equal intermediate lifetimes leading to a double peak in the ejected mass distribution at these intermediate delay times.  

The probability to survive to age $t$ and then make the supernova phase transition
in the next interval $dt$ is
\be
    dP(t) = P_s(t) \, dt/\tau(M(t))\quad .
\ee
The energy released in the phase transition is significantly greater than the 
energy released afterwards by carbon fusion which results in the star being totally
disrupted except for a very small remnant discussed in Section \,\ref{remnants}.
The supernova double distribution as a function of ejected mass, $M(t)$ and delay time $t$ is therefore,
\be
    \frac{d^2 N_{SN}}{dM(0) dt} = a_0\, F(M(0))\,P_s(t)/\tau(M(t))  
\label{doubledist}
\ee
with
\be
    M(t) = M(0) +c_0 t\quad .
\ee

The situation with respect to 
delay times is well summarized in the graph of ref\,\cite{Maoz12} reproduced 
here in fig.\,\ref{DTdist}.
The recovered delay times \cite{Maoz12} with the smallest statistical errors 
are shown in filled red circles in this graph.  Previous measurements \cite{M11} with larger statistical errors are shown in open blue squares shifted slightly to the left for clarity.  Measurements from 2010 \cite{Maoz2010} are shown in open black circles.   These latter, unfortunately, are in statistical disagreement with the later measurements implying one or more methods are subject to biases or large systematic errors.

\begin{table}[ht]
\begin{center}
\begin{tabular}{|l|c|c|c|}\hline
      &\,$  \mathrm{min}  $  &\; $  \mathrm{variable}$\;  &\, $ \mathrm{max} $\cr\hline
\;$\mathrm{scan:}\,$ &\,$ 5\cdot 10^7\,\mathrm{g/cm^3} $&\,$ \rho_c $ &\, $1.5\cdot 10^8\,\mathrm{g/cm^3} $\cr
      &\,$ 0.15\,\mathrm{Gyr} $ &\,  $ \tau_0 $  &\,$ 0.45\,\mathrm{Gyr}$ \cr
      &\, 0     &\, $  b_0  $ &\, 4\cr
      &\, 0     &\,  $ c_0  $ &\, $0.01\,M_\odot/\mathrm{Gyr}$ \cr \hline      
\;$\mathrm{require:}\,$ &\, $1.11\cdot 10^{-2} $&\,  $ S_1 $  &\,$ 1.71 \cdot 10^{-2} $\cr
         &\, $ 0.135 $ &\, $   R_{21} $   &\,$  0.227 $ \cr \hline
\;$\mathrm{find:}\,$ &\; $ 7.3\cdot 10^7\,\mathrm{g/cm^3} $&\, $  \rho_c $  &\, $1.45\cdot 10^8\,\mathrm{g/cm^3}\;$  \cr
      &\,  0.18  &\, $  \tau_0   $ &\,  0.39  \cr
      &\,  0.16  &\,   $b_0$    &\,  4  \cr
      &\, 0 &\,   $c_0$    &\, $ 0.01\,M_\odot /\mathrm{Gyr}$ \cr
      &\,  $1.7 \cdot 10^{-3} $&\,  $ S_2  $ &\, $3.2\cdot 10^{-3} $ \cr
      &\,  $1.1 \cdot 10^{-4} $&\,  $ S_3  $ &\, $9.3\cdot 10^{-4} $ \cr
      &\,  0.009 &\, $  R_{SN}  $ &\, 0.023  \cr\hline
\end{tabular}
\end{center}
\caption{Four parameter monte carlo in phase transition model showing the 
range of scanned variables, the conditions required, and the resulting minimum
and maximum values of the output parameters.}
\label{table1}
\end{table}
\begin{figure}[ht]
\centering
\includegraphics[scale=0.65]{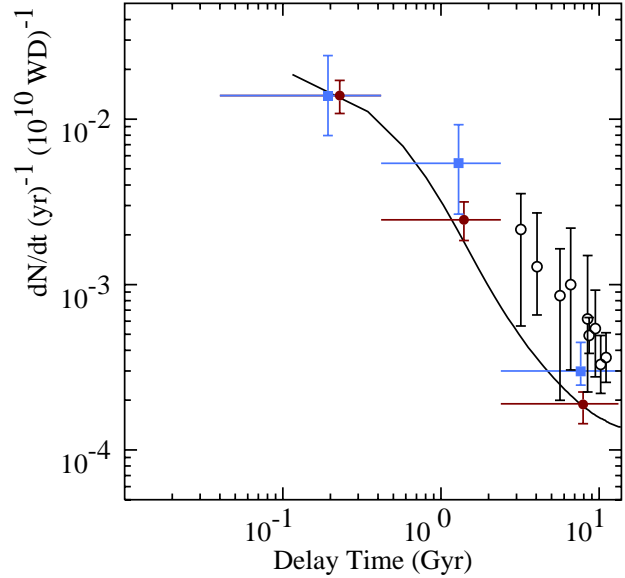}
\caption{The SNIa rate as a function of delay time in the phase transition model
with critical density $\rho_c = 9\cdot 10^7 \mathrm{g/cm^3}, \tau_0 = 0.326
\mathrm{Gyr}, b_0 = 3.4$, and $c_0 = 0.0016 M_\odot/\mathrm{Gyr}$.  Data is taken  
from the graph of ref.\,\cite{Maoz12} reprinted here in fig.\ref{DTdist} .} 
\label{DTdist2}
\end{figure}

In refs.\,\cite{M11} and \cite{Maoz12} the supernova rates were divided into three delay time bins bounded by the delay times:
\be
    t_1=0.04 Gyr \quad t_2 = 0.4 Gyr \quad t_3 = 2.4 Gyr \quad t_4 = 13.8 Gyr \quad .
\ee
The average supernova rate in the three bins is 
\be
     S_i = \frac{1}{t_{i+1}-t_i}\,\int_{t_i}^{t_{i+1}} dt\,\frac{d N_{SN}}{dt}
\quad .
\ee

In the phase transition model, the integral over delay times at fixed
initial mass can be done analytically so that 
\be
  S_i =  \frac{1}{t_{i+1}-t_i}\,\int dM(0) \int_{t_i}^{t_{i+1}} \frac{d^2 N_{SN}}{dM(0) dt} dt = a_0\,\int dM(0) F(M(0)) \frac{P_s(t_{i}) -P_s(t_{i+1})}{t_{i+1}-t_i}   \quad .
\label{Si}
\ee
The total supernova Ia rate is
\be
    R_{SN} = a_0\,\int dM(0) F(M(0)) (P_s(0) -P_s(t_4))  \quad .
\label{totalRate}
\ee

\subsection{The DD scenario\label{subsec:DD}}{
The approximate linearity of the curve joining the filled red points in fig.\ref{DTdist}
with a power of approximately $-1$ ($p = -1.12 \pm 0.08$)
has been interpreted as supporting the DD scenario providing the initial separation distribution of the two white dwarfs is approximately $t^{-1}$. 
One can
question whether this fit requires an implausibly high frequency of binary white
dwarfs with a high combined mass.  
For instance, if we assume in the DD scenario
that a collapsing star that would lead to a mass between 1.35 and 1.45 produces with probability $q$ a binary white
dwarf system with that total mass, the birthrate (eq.\ref{birthrate}) of such stars and an upper limit to the supernova rate should, from the Salpeter initial mass function, be
\be
    dN/dt = a_0 \,q\,\int_{1.35}^{1.45} F(M(0)) dM(0) = q \, 1.8 \cdot 10^{-3} 
\mathrm{SN/yr/gal} \quad .
\ee
Since $q$ must be no more than unity this underpredicts the SNIa rate by a factor of about $10$ as is confirmed in more
detailed treatments.  The alternative path to a binary initial state of white dwarfs
namely independent but simultaneous production of white dwarfs from nearby main sequence stars is probably no more likely.  In the clean sample of ref.\cite{Bergeron}
only $13\%$ of white dwarfs are known or suspected to be in double degenerate configurations and, of these, the heaviest dwarf has mass of only $0.66$. 

Moreover, with respect to the linear fit,
the required inverse $d$ initial separation distribution is not strongly motivated from theory, finiteness requires that the linearity fails at small delay times, and
in addition the filled red data points might show a slight negative curvature, i.e. the observed value of $S_2$ is some two standard deviations higher than the best fit.
As we will see the delay time distribution can be adequately fit in the phase transition model.   
}
Returning to the consideration of the phase transition model,
from the graph of fig.\,\ref{DTdist} and the data of ref.\,\cite{Maoz12} we read the
$1\sigma$ SNIa rates in the three bins in units of SNIa/yr per prototype galaxy
of $10^{10}$ white dwarfs:
\be\nonumber
  1.11 \cdot 10^{-2} < S_1 < 1.71 \cdot 10^{-2} \\\nonumber
  1.88 \cdot 10^{-3} < S_2 < 3.16 \cdot 10^{-3} \\
  1.44 \cdot 10^{-4} < S_3 < 2.33 \cdot 10^{-4}
\label{reddata}
\ee  
or
\be\nonumber
   0.135 < R_{21} \equiv S_2/S_1 < 0.227 \\
   0.008 < R_{31} \equiv S_3/S_1 < 0.021 
\ee 

An advantage of the ratios $R_{21}$ and $R_{31}$ is that they are independent of the
$a_0$ parameter, i.e. they should be the same for any sample of white dwarfs
following the high mass part of the hot white dwarf distribution in 
fig.\,\ref{hotDAdwarfs} regardless of the total number of white dwarfs in the sample.  
To the extent that the DB dwarfs have the same initial mass function and a 
similar accretion rate their effect is included in the predicted supernova
rates per $10^{10}$ white dwarfs.  The phase transition probability would not
differ between dwarfs with a thin atmosphere of hydrogen or helium.
In the phase transition monte carlo we scan over the four parameters requiring
that the resulting $S_1$ and $R_{21}$ fall within their $1 \sigma$ ranges.
The resulting ranges of the other quantities are then tabulated in table\,\ref{table1}.

The values for the basic parameters, $\rho_c$ and $\tau_0$ are not far
from the estimates of ref.\,\cite{Biermann-Clavelli} made prior to the
latest data on delay times and ejected mass.  Fixing only $S_1$ and $R_{21}$
the model predictions span the range of observations for $S_3$
and the total supernova rate.
In the phase transition model the supernovae fall naturally into two
classes, those occuring in isolated dwarfs and those occuring in
binary systems with significant accretion onto the white dwarf. 
The first class contains only those supernovae initiated by the vertical
ageing in fig.\,\ref{ageing} which would be expected to have a lower $S_3$.
The second class is represented by the curved paths in fig.\,\ref{ageing}
and would have higher $S_3$ values. If the delay time recovery method is 
not biased
the delay time distribution will be an appropriately weighted average of the
two classes.  In Table\,\ref{table2} we separate the results in the two classes.

\begin{table}[ht]
\begin{center}
\begin{tabular}{|l||c||c|}\hline
\multicolumn{1}{|c||}{require:}&
\multicolumn{2}{|c|}{ $0.011 < S_1 < 0.0171$}\\
\multicolumn{1}{|c||}{}&
\multicolumn{2}{|c|}{$ 0.135 < R_{21} < 0.226$}\\\hline
\multicolumn{1}{|c||}{scan:}&
\multicolumn{2}{|c|}{$ 5\cdot 10^7 \mathrm{g/cm^3}< \rho_c < 1.5\cdot 10^8
\mathrm{g/cm^3}$}\\
\multicolumn{1}{|c||}{}&
\multicolumn{2}{|c|}{$ 0.15 \mathrm{Gyr} < \tau_0 < 0.45 \mathrm{Gyr}$}\\
\multicolumn{1}{|c||}{}&
\multicolumn{2}{|c|}{$ 0 < b_0 < 4$}\\
\multicolumn{1}{|c||}{}&
\multicolumn{1}{|c||}{$0 < c_0 < 0.002\,M_\odot/\mathrm{Gyr}$}&
\multicolumn{1}{|c|}{$0.002 M_\odot/\mathrm{Gyr}<c_0<0.01\, M_\odot/\mathrm{Gyr}$}\\
\hline
\multicolumn{1}{|c||}{find:}&
\multicolumn{1}{|c||}{$0.246 \,\mathrm{Gyr} < \tau_0 <0.389 \,\mathrm{Gyr}$}&
\multicolumn{1}{|c|}{$0.184\, \mathrm{Gyr} < \tau_0 <0.389 \,\mathrm{Gyr}$}\\\cline{2-3}
\multicolumn{1}{|c||}{}&
\multicolumn{1}{|c||}{$1.7\cdot 10^{-3}  < S_2 < 2.9\cdot 10^{-3}$}&
\multicolumn{1}{|c|}{$1.7\cdot 10^{-3}  < S_2 < 3.2\cdot 10^{-3}$}\\\cline{2-3}
\multicolumn{1}{|c||}{}&
\multicolumn{1}{|c||}{$1.1\cdot 10^{-4}  < S_3 < 2.6\cdot 10^{-4}$}&
\multicolumn{1}{|c|}{$1.9\cdot 10^{-4}  < S_3 < 9.3\cdot 10^{-4}$}\\\cline{2-3}
\multicolumn{1}{|c||}{}&
\multicolumn{1}{|c||}{$0.0068  < R_{31} < 0.0187 $}&
\multicolumn{1}{|c|}{$0.014  < R_{31} < 0.091 $}\\\hline
\end{tabular}
\end{center}
\caption{Four parameter monte carlo in phase transition model
separated into the classes with small or large accretion rates. As seen here
a larger accretion rate leads to a larger $S_3$ as suggested in the data of ref.\,\cite{Maoz2010}. Output variables that are
insensitive to the accretion rates are not shown. }
\label{table2}
\end{table}

The parameter space of the phase transition model could be further restricted if
the tension between the high delay time measurements could be resolved.  
If the lower $S_3$ values are confirmed, the accretion rates favored by the monte carlo are only about $1\%$ of those allowed by the x-ray data 
\cite{Bogdan-Gilfanov}
which are themselves only about $1\%$ of the rates that would be required if the 
full supernova Ia rate were to be explained in the single degenerate scenario.
The prediction of low average accretion rates is significant since, with larger
accretion rates, the absence of binary partner effects would be puzzling even
in the phase transition model.

\section{The ejected mass distribution\label{massdists}}
 
\begin{figure}[ht]
\centering
\includegraphics[scale=0.65]{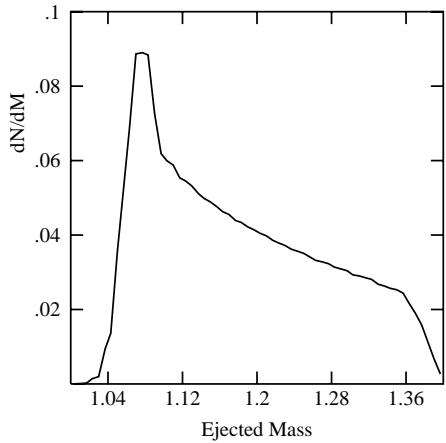}
\caption{The ejected mass distribution for $\rho_c = 9\cdot 10^7\,\mathrm{g/cm^3},\, \tau_0=0.326\,\mathrm{Gyr}\,, b_0 = 3.42,\, \mathrm{and}\, c_0 = 0.0016\, M_\odot/\mathrm{Gyr}$} 
\label{EjectedMass}
\end{figure}

Attempts to discover remnants of SNIa have up to now been unsuccessful suggesting
the supernova process totally disrupts the white dwarf.  This in turn implies 
in standard scenarios that the ejected mass should be very close to the Chandrasekhar mass.  However, recent studies have shown ejected mass distributions
extending down to $0.9\,M_\odot$\,\cite{Scalzo}.  That would imply remnants extending up to $0.5\,M_\odot$.  Numerous white dwarfs have been discovered in this mass range but never as supernova remnants.  As noted in that reference the ejected mass distribution strongly suggests a sub-Chandrasekhar mechanism such as supplied by the present phase transition model.

From eq.\,\ref{doubledist} we can write the ejected mass distribution between delay times of $t_i$ and $t_{i+1}$ assuming the
supernova totally disrupts the white dwarf as suggested by the absence of
significant remnants and predicted in the present model.
\be
     \frac{dN}{dM} = a_0 \int_{t_i}^{t_{i+1}}\,dt\,F(M - c_0 t)\,P_s(t)/\tau(M)
\quad .
\ee
Integrating over all delay times, the full ejected mass distribution is
\be
     \frac{dN}{dM} = a_0 \int_{0}^{13.8 \mathrm{Gyr}}\,dt\,F(M - c_0 t)\,P_s(t)/\tau(M)
\quad .
\ee

Neglecting $c_0$, the time integral here can be done analytically.
The numerical evaluation of the full ejected mass distribution is shown in fig.\,\ref{EjectedMass}
for a particular choice of model parameters.  The peak of the ejected mass
distribution is somewhat lower than indicated in the data of ref.\,\cite{Scalzo}
but its narrowness is interesting from the point of view of
supernova homogeneity.  The peak would move to the right if the critical
density parameter were increased.

\begin{figure}[ht]
\centering
\includegraphics[scale=0.65]{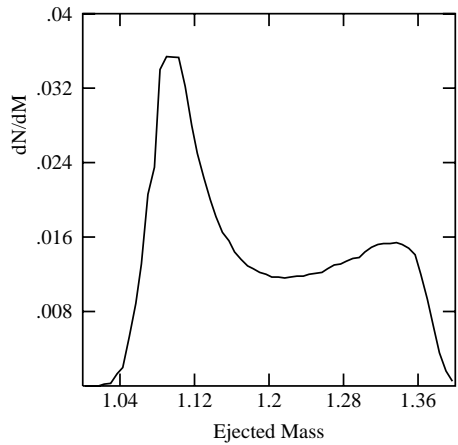}
\caption{The ejected mass distribution for the model parameters used in
fig.\,\ref{EjectedMass} but restricted to the intermediate delay times
($0.4\, \mathrm{Gyr} <t<2.4\,\mathrm{Gyr}$).}  
\label{S2EjectedMass}
\end{figure}

As pointed out in section\,\ref{Carlo}, at intermediate delay times the phase
transition model predicts a double peak in the ejected mass distribution.  
This is illustrated in fig.\,\ref{S2EjectedMass}.

One should note that the present model for sub-Chandrasekhar supernovae does
not preclude a few percent of supernovae coming from accretion to the Chandrsekhar mass as is consistent with the x-ray data \cite{Bogdan-Gilfanov}.  These events
arise from white dwarfs punching through the low lifetime region due to
rapid accretion.  The result would be a second peak in the ejected mass distribution as perhaps observed in the data of ref.\,\cite{Scalzo}.
This is independent of the second peak expected in the current model
at intermediate delay times as mentioned in section\,\ref{Carlo}.

\section{SNIa remnant in the Phase Transition Model\label{remnants}}

One of the great puzzles in the standard model of SNIa is the absence of detected remnants.
In the phase transition model a small bubble of exact susy grows releasing
Pauli energy until that energy plus the energy released through fusion
is equal at least to the energy required to unbind the shell outside.  
If one treats
the white dwarf as having the average density $3 M/(4 \pi R^3)$ the remnant
has about $10^{-3}$ of the initial mass \cite{Biermann-Clavelli}.
In a more precise treatment of the rapidly decreasing density with radius
one might expect this $10^{-3}$ fraction to decrease.    
This remnant is predominantly made of the scalar partners of the protons,
neutrons, and electrons.  Since these cannot be supported by degeneracy 
pressure, the remnant collapses to a black hole of approximately
Jupiter mass or smaller.  Such a small black hole would be difficult to detect in the supernova aftermath. In the typical galaxy there have been about
$10^{8}$ type Ia supernovae since the big bang so the phase transition 
model would predict this
number of low mass black holes in the Milky Way.  Charged particles
falling into a black hole could be expected to radiate a few percent of their
rest mass.  This prediction
could perhaps be tested in cosmic x-ray backgrounds \cite{SS}\,,\cite{Moran}\,,\cite{Moretti}.  In addition many stars
could have captured one of these small black holes which might then be
manifested by massive objects orbiting at large angle from the solar disks
similar to the apparent ``planet nine" in our own solar system.  Such black
hole planets might also be expected to have their own moons and/or
accretion disks.  The complete treatment of the growth of a susy bubble in 
a dense star and its remnant is a complicated dynamical calculation which has 
only been preliminarily considered in refs.\,\cite{Clav-Perevalova},\cite{growth},\cite{gravcollapse}.

\section{Summary\label{summary}}

The phase transition model has a number of advantages relative to the 
standard model explosion at the Chandrasekhar mass.  The delay time distribution
although not linear in a log-log plot can easily fit the data.  The
ejected mass distribution naturally peaks below the Chandrasekhar 
mass. The total supernova rate is related to known single white dwarf mass
distributions unlike the situation in the standard model which depends on unknown binary distributions and which, with plausible assumptions, greatly underestimates the total rate.  The absence of observed remnants which is puzzling in the 
standard model
is easily understood in the phase transition model which also makes
an interesting prediction of remnant effects. The sphericity of supernova
events and the absence of shadowing by a binary partner are predicted in the phase transition model with low accretion rates but remain puzzling in the standard model. In the DD scenario one must wonder why there are no observations of
binary white dwarf systems with near Chandrasekhar combined mass and how the
homogeneity of normal SNIa which is crucial to dark energy measurements is maintained.
In the DD scenario one would expect that the peak of the white dwarf distribution at $0.6$
would lead to a secondary peak at $1.2$ due to a coalescence of binary white dwarfs.
Indeed in fig.\,\ref{hotDAdwarfs} there seems to be such a secondary peak but it is known 
\cite{Madej04} that this is purely an artifact of the treatment of the high mass tail.

Significant discussions with Peter Biermann, Ken Olum, and Akos Bogdan 
contributing to the work of this paper are gratefully acknowledged.

\end{document}